\documentclass[twocolumn,superscriptaddress,showpacs,preprintnumbers,amsmath,amssymb]{revtex4}
\usepackage{graphicx}
\usepackage{dcolumn}
\usepackage{bm}
\begin{document}
\title{Geometric quantum computation and multi-qubit entanglement
with superconducting qubits inside a cavity}
\author{Shi-Liang Zhu}
\affiliation{School of Physics and
Telecommunication Engineering, South China Normal University,
Guangzhou, China} \affiliation{Institute for Scientific
Interchange Foundation, Viale Settimio Severo 65, I-10133 Torino,
Italy}
\author{Z. D. Wang}
\affiliation{Department of Physics, University of Hong Kong,
Pokfulam Road, Hong Kong, China} \affiliation{National Laboratory
of Solid State Microstructures, Nanjing University, Nanjing,
China}
\author{Paolo Zanardi}
\affiliation{Institute for Scientific Interchange Foundation,
Viale Settimio Severo 65, I-10133 Torino, Italy}

\begin{abstract}
We analyze  a new scheme for quantum information processing, with
superconducting charge qubits coupled through a cavity mode, in
which  quantum manipulations are  insensitive to the state of the
cavity. We illustrate how to physically implement universal
quantum computation as well as multi-qubit entanglement based on
unconventional geometric phase shifts in this scalable solid-state
system.  Some quantum error-correcting codes can also be easily
constructed using the same technique. In view of the gate
dependence on just  global geometric features and the
insensitivity to the state of  cavity modes, the proposed quantum
operations may result in   high-fidelity  quantum information
processing.

\end{abstract}

\maketitle

\newpage

Superconducting qubits have recently attracted significant
interests because of their potential suitability for integrated
devices in quantum information
processing\cite{Makhlin,Nakamura,Yu,Vion}. So far,  experimental
research on quantum information processing using this kind of
systems has mostly focused on the behavior of single isolated
qubit as the decoherence time of this solid-state system is quite
short\cite{Makhlin}. Only very recently, significant achievements
on superconducting two-charge-qubit systems were reported, i.e.,
realization of an entangled state for two-qubits, and
implementation of a conditional gate \cite{Nakamura}. Also note
that multiparticle entanglement was experimentally reported only
in photons and trapped ions. Implementation of a universal set of
high-fidelity quantum gates and generation of multi-qubit
entanglement will be the next significant and very challenging
steps towards quantum information processing based on this
scalable solid-state approach.

In this paper, by designing a  device consisting of
superconducting charge qubits coupled through a cavity, we propose
a feasible scheme to implement a universal set of quantum gates
and produce the Greenberger-Horne-Zeilinger state (GHZ)\cite{GHZ}
based on unconventional geometric phases.
In addition, we
find that this scheme is also  applicable to construct quantum
error-correcting codes (QECC)\cite{Shor}. In particular, we shall
show  how to realize the  geometric evolution operators $U_{x,y}
(\gamma)= \exp(i\gamma J^2_{x,y})$ with $J_{x,y}$ collective
operators and $\gamma$ an unconventional geometric
phase\cite{Zhu_PRL2003}, which may have some inherent
fault-tolerant features   due to the fact that unconventional
geometric phases depend only on some global geometric property
\cite{Berry,Zanardi,Duan,Falci,Zhu_PRL2002,Leibfried,Zhu_PRL2003}.
It is also shown that in terms of $U_{x,y} (\gamma)$ of the
present system one is able to achieve a universal set of quantum
gates, to generate even-N-qubit entanglement simultaneously by one
operation\cite{Sorensen2000}, and to construct the QECC\cite{You}.
All of these  are essential ingredients in quantum information
processing. Implementation of these tasks based on the same set of
geometric quantum operators may simplify the experimental
operations. Apart from the high-fidelity advantage, the geometric
scheme proposed here has other distinctive merits: the operator is
insensitive to the state of the cavity modes
\cite{Sorensen2000,Milburn}, is tolerant to device parameter
nonuniformity in quantum computation, and the process can be
fast\cite{Zheng,Garcia-Ripoll}. Moreover, a necessary condition
for fault tolerant quantum computation, i.e., the two-qubit gate
can act on any pair of qubits, can be realized using the cavity
quantum electrodynamic technique. In addition,  the system
proposed is also a potential candidate for quantum communication
and quantum network.

  Superconducting nanocircuits coupling through cavities
have been shown to be a promising solid-state system for
implementation of quantum computation and quantum
communication\cite{Zhu_WY,Yang,Girvin,Wallraff}. One of advantages
for qubits placed in a cavity is that the cavity can also protect
the qubits from the environment, which is important for a useful
operation of qubits especially in the scaling-up of the
solid-state devices. Moreover, an architecture using
one-dimensional transmission line resonators to reach the strong
coupling limit between cavity and superconducting nanocircuits was
theoretically proposed in Ref.\cite{Girvin} and then
experimentally achieved in Ref.\cite{Wallraff}. In the present
work, we address a system with $N$ specially designed
superconducting charge qubits coupled through a high-quality
single mode cavity, as shown in Fig. 1. Clearly, as shown in
Fig.1(a), the newly designed single qubit is significantly
different from those studied
before\cite{Makhlin,Nakamura,Falci,Zhu_WY}. The $j$th qubit
consists of a small superconducting box with $n_j$ excess
Cooper-pair charges, formed by two SQUIDs with Josephson coupling
energies $E_j^{\epsilon l}$ $(\epsilon=a,b; l=1,2)$ and a
$\pi$-phase junction, and each SQUID is penetrated by a magnetic
flux $\phi_j^\epsilon$. A control gate voltage $V_j^g$ is
connected to the system via a capacitor $C_j^g$. The Hamiltonian
of the $j$th qubit reads
\begin{equation}  \label{H_single}
H=E_{ch} (n_j-\bar{n}_j)^2-\sum_{\epsilon l }E_{j}^{\epsilon l}
\cos\varphi_j^{\epsilon l},
\end{equation}
where
$E_{ch}=2e^2/C_j$ is the charging energy with $C_j$ being the
total capacitance of the box, $\bar{n}_j=C_j^g V_j^g/2$ is the
induced charge and can be controlled by changing $V_j^g$. The
gauge-invariant phase difference $\varphi^{\epsilon
l}_j=\Theta_j-(2\pi/\phi_0)\int {\bf A}^{\epsilon l}_j\cdot d{\bf
l}$ with $\Theta_j$ being the phase difference of the
superconducting wave function across the junctions in a particular
gauge and ${{\bf A}}^{\epsilon l}_j$ being the vector potential in
the same gauge. Assuming that Josephson junctions are placed
inside a single-mode resonant cavity, we have $(2\pi/\phi_0)\int
{\bf A}_j^{\epsilon l}\cdot d {\bf l} =(2\pi/\phi_0) \int {{\bf
A^\prime}_j^{\epsilon l}} \cdot d {\bf l} + g_j(a+a^{\dagger})$,
where $a$ and $a^\dagger$ are the creation and annihilation
operators for the single-mode,  $\phi_0=\pi\hbar/e$ is the flux
quantum, $g_j$ is the coupling constant between the junctions and
the cavity, and ${{\bf A^\prime}_j^{\epsilon l}}$ is the vector
potential determined by the magnetic flux $\phi^\epsilon_j$
penetrating the $\epsilon$th SQUID hole in the $j$th qubit.

\begin{figure}[htbp]
\vspace{-0.5cm} \centering
\includegraphics[height=6cm,width=8cm]{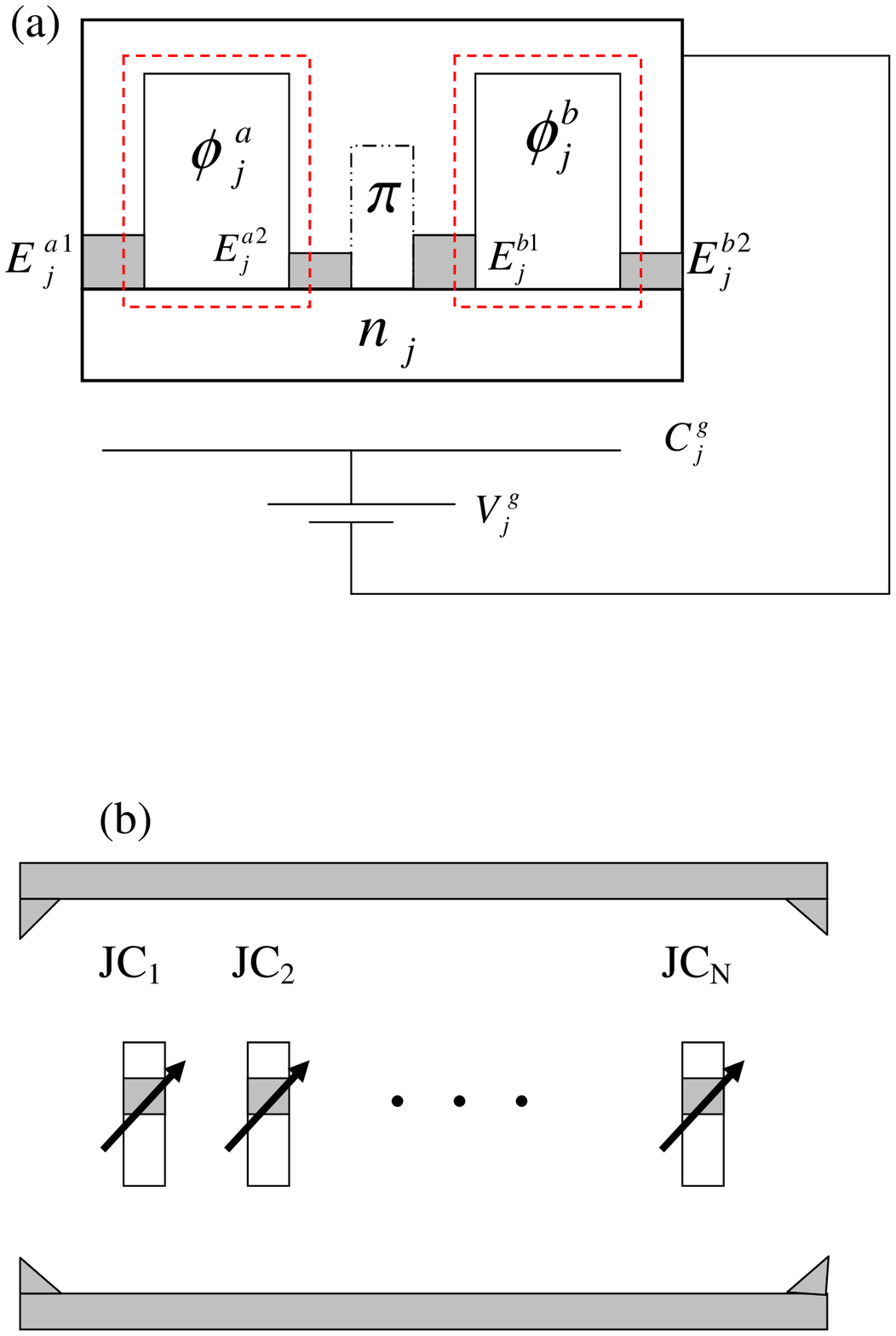}
\vspace{-1.0cm}
 \caption{Josephson charge qubits. (a) A single
Josephson charge qubit. (b) A series of Josephson charge qubits
coupled through a cavity} \label{fig1}
\end{figure}

At temperatures much lower than the charging energy and the gate
voltage being tuned close to a degeneracy ($\bar{n}_j\sim 1/2$),
the relevant physics is captured by considering only the two
charge eigenstates $n_j=0,1$, which constitute the basis
$\{|0\rangle,|1\rangle\}$ of the computational space of the qubit.
If we have $N$ such qubits located inside a single-mode cavity
[Fig.1(b)] with frequency $\omega_c/2\pi$, to a good
approximation, the whole system can be considered as $N$ two-state
systems coupled to a quantum harmonic oscillator. On the other
hand, $\varphi_j^{\epsilon l}$ are determined from the flux
quantization for two independent loops and an effective
$\pi$-phase junction between $a2$ and $b1$ for each qubit [see Fig
1(a)], that is, $\varphi_j^{a1}-\varphi_j^{a2}=\Phi_j^a$,
$\varphi_j^{b1}-\varphi_j^{b2}=\Phi_j^b$, and
$\varphi_j^{a2}-\varphi_j^{b1}=\pi$\cite{Pi_flux},
 where the flux
$\Phi_j^\epsilon=\tilde{\phi_j}^\epsilon + g_j (a+a^\dagger)$ with
$\tilde{\phi_j}^\epsilon=2\pi\phi_j^\epsilon/\phi_0 $. Moreover,
the average phase drop $\sum_{\epsilon l} \varphi_j^{\epsilon
l}/4$ is equal to $\Theta_j$, which conjugates to the Cooper pair
number $n_j$. Here we assume that $E_j^{al}=E_j^{bl}=E^{l}_j$ and
$g_j^\epsilon=g_j$.
In this case, the whole system can
be described by the Hamiltonian $H = H_0+H_{int}^\prime$, where
\begin{equation}
\label{H_0} H_0 = \hbar \omega_c (a^\dagger a+\frac{1}{2}) +
\sum_{j}^{N} E_{\bar{n}_j} \sigma_j^z/2,
\end{equation}
\begin{equation}
\label{H_int1} H_{int}^\prime = -\frac{1}{2}\sum_{l j} [i (-1)^l
E^{l}_j  (1- e^{i\Gamma_j^l})e^{-i\phi_j^{-}/2} \sigma^{+}+H.c. ].
\end{equation}
Here $E_{\bar{n}_j}=2E_{ch}(\bar{n}_j-1/2)$,
$\Gamma_j^l=\phi_j^{-}+ i (-1)^{l-1} [\phi_j^{+} + g
(a+a^\dagger)]$, and $\phi_j^{\pm}=(\tilde{\phi}_j^a \pm
\tilde{\phi}_j^b)/2$. A spin notation is used for the qubit $j$
with Pauli matrices $\{ \sigma^{x}_j,\sigma^{y}_j,\sigma^{z}_j
\}$, and $\sigma_j^{\pm}=(\sigma^x_j \pm i\sigma^y_j)/2$.
It is interesting to note that this Hamiltonian is similar to that
used  in Ref.\cite{Sorensen2000} when the external magnetic flux
$\tilde{\phi}_j^\epsilon$ depends on the time $t$ with a constant
rate $\omega_j^\epsilon $.

Comparing with the charge qubit made of one SQUID and coupled to
the cavity\cite{Zhu_WY}, where the initial cavity mode should be
in the ground state in performing quantum gates(the task is
exceedingly difficult to implement experimentally), one more SQUID
being penetrated through another  magnetic flux is used here. Two
magnetic fluxes for one qubit can be used to manipulate the qubit
states. The function of these fluxes is similar to bichromatic
lights in quantum computation based on trapped
ions\cite{Sorensen2000}: transition paths involving unpopulated
cavity states interfere destructively to eliminate the influence
of the cavity mode. This phenomenon
 plays a key role in the implementation of a quantum gate which is insensitive
to the initial state of cavity mode and robust against changes in the
cavity state occurring during operation. This scenario has been
 used in  implementation of quantum computation with trapped
ions\cite{Leibfried,Sorensen2000,Zheng,Milburn,XWang,Sackett,Garcia-Ripoll},
and in generating experimentally  the maximally entangled state
for four ions\cite{Sackett}. In the following we illustrate how
this strategy  can be applied to  the present system.

 To eliminate the influence of the cavity mode, we now
 propose to apply two constantly growing external
fluxes penetrating the designated SQUIDs in the $j$-th qubit:
 $\tilde{\phi}_j^{a} =\omega_j^\phi t-\beta^0_j$ and $
\tilde{\phi}_j^{b} =\omega_j^\phi t - \beta^0_j - 2 k \pi$,  where
$\beta^0_j$ ( $\beta_j^0 - 2k \pi$) is the initial value of
$\tilde{\phi}_j^a$ $(\tilde{\phi}_j^b)$  and $k$ is an integer.
Experimentally, one possible way is to place each SQUID  just
above an approximately-inductive circuit
with a constant voltage $V_0=\omega_j^\phi \phi_0/2\pi$.
Expanding
the Hamiltonian (\ref{H_int1}) to the first order of $g_j$ in the
Lamb-Dicke limit and under the rotating wave approximation as well
as in the interaction picture $U_0=\exp(-iH_0 t)$ with
$E_{\bar{n}_j}=0$ (i.e., the qubits are set at the degeneracy
point) and $E_j^1=E_j^2=E_j$ , yields
 \begin{equation}
\label{H_eff1} H_{int}^\eta=\sum_j^N g_j E_j( a^\dagger
e^{i\delta_j t+i\beta_j^0}+H.c.) \sigma_j^\eta ,\ \ (\eta=x,y)
\end{equation}
 where $\eta=x,y$ for  $k=0, 1$ , and $\delta_j=\omega_c-\omega_j^\phi
<<\omega_j^\phi$ \cite{Note}.

We now show that the Hamiltonian described in Eq.(\ref{H_eff1})
can be used to perform universal  quantum computation as well as
to produce the GHZ state. Let us first address a  simpler case, in
which all parameters for different qubits are the same, namely,
$E^{\epsilon l}_j=E$, $\delta_j=\delta$, and $\beta_j^0=\beta^0$.
The corresponding Hamiltonian (\ref{H_eff1}) can be rewritten as

\begin{equation}
H_{int}^\eta=g E( a^\dagger e^{i\delta t+i\beta^0}+H.c.)J_\eta,
\end{equation}
where $J_\eta = \sum_j^N \sigma_j^\eta$ is the collective spin
operator. By using the magnus' formula, the evolution operator
$U(t)$ is found as\cite{Zhu_PRL2003}
\begin{equation}
\label{U_mode} U_\eta(t)=e^{i\gamma (t) J_\eta^2} e^{[\alpha
(t,\beta^0) a^\dagger-\alpha^\star (t,\beta^0) a ] J_\eta},
\end{equation}
where $\alpha (t,\beta^0)=(gE/\hbar \delta)(1-e^{i\delta
t})e^{i\beta^0}$ and $\gamma (t)=(gE/\hbar\delta)^2 (\delta t-
\sin\delta t)$.
Here $\alpha(t,\beta^0)$ is a periodic function of time and
vanishes at $t_m=2 m\pi/\delta$ for an integer $m$. At the time
$t=t_m$, the evolution operator is explicitly expressed as
\begin{equation}
\label{U_x} U_\eta(\gamma)=\exp[i\gamma  J_\eta^2].
\end{equation}
This operator is insensitive to heating by removing the influence
of the cavity mode represented by the last exponent in
Eq.(\ref{U_mode}).

 More intriguingly, we can remove the influence of the cavity mode by using two
operators in succession\cite{Zheng}. Let us suppose that qubits
first evolve with $\beta^0=0$ for a period $\tau/2$. Then
 the $\beta^0$ is shifted by $\pi$ for the other period $\tau/2$. Note
that $\alpha (t,0)=-\alpha (t,\pi)$ and $\gamma (t)$ is
independent on $\beta^0$, we have $ U_\eta=\exp[i 2\gamma(\tau/2)
J_\eta^2]$. Comparing with the first approach, a distinctive merit
lies in that   the gate operation time is no longer restricted to
be
 $2 m\pi/\delta$ \cite{Zheng} and it can then be shorter;
 of course a gate with fast speed is
particularly important in solid state qubit systems since the
decoherence time there is typically quite short.

Interestingly, the evolution operators (\ref{U_x}) can provide  a
set of universal quantum gates. It is well known that for
achieving universal quantum computation, we need to realize  two
noncommuting single-qubit gates and one nontrivial two-qubit gate.
When only two qubits are considered, it is straightforward to
check that $U_x$ (or $U_y$) is a nontrivial two-qubit gate. $U_x$
can be used to produce an entangled state from an untangled state.
For example, the maximally entangled state
$(|00\rangle-i|11\rangle)/\sqrt{2}$ is derived when $U(\pi/2)$ is
directly applied to the untangled state $|00\rangle$. Moreover, a
controlled-NOT gate is explicitly generated by $U_x$ plus
single-qubit rotations in Ref.\cite{Sorensen2000}. We now work out
how to use the operators $U_{x,y}$ to achieve a set of
noncommuting single-qubit rotations if one of qubits is an
auxiliary qubit. Denoting $|\psi_{\pm}^{x,y}\rangle_a$ as the
eigenstate of the operator $\sigma_{x,y}$ for the auxiliary qubit
with eigenvalue $\pm1$, and choosing the initial state as one of
the eigenstates, then we have
\begin{equation}
\label{Single}
U_{x,y}|\psi\rangle|\psi_{\pm}^{x,y}\rangle_a=e^{2i\gamma}[e^{\pm
2i\gamma\sigma_{x,y}}|\psi\rangle]|\psi_{\pm}^{x,y}\rangle_a.
\end{equation}
It is clear from Eq.(\ref{Single}) that an effective single qubit
gate $U_{x,y}^{(1)}=\exp(\pm 2i\gamma\sigma_{x,y})$ is obtained
(up to an irrelevant overall phase) by application of $U_{x,y}$ to
two qubits, but one of them is an auxiliary qubit with a fixed
initial state and then is disregarded after the gate operation.
Since $|\psi_{\pm}^{x}\rangle_a$ and $|\psi_{\pm}^{y}\rangle_a$
are the ground states of the Hamiltonian at
($\bar{n}_a=1/2,\phi^{-}=0)$ and ($\bar{n}_a=1/2,\phi^{-}=\pi)$,
respectively, it is rather easy to experimentally realize the
required initial state in the present system. Certainly,
$U_x^{(1)}$ and $U_y^{(1)}$ are noncommuting and consist of the
well-known single-qubit rotations. Although one more auxiliary
qubit is required, all important operations required (all logical
gates and construction of QECC addressed below) can be achieved
with similar manipulations, which may simplify experimental
operations. In addition, comparing with qubits coupled through
capacitance, a nearest neighbor interaction, the distinctive merit
is that the nontrivial two-qubit gate proposed here may act on any
pair of qubits.

It is worth pointing out that a non-trivial two-qubit gate can
also be implemented when the parameters $g_j$ and $E_j$  are
dependent on $j$, thus the method is also insensitive to
fabrication errors often appeared in realistic solid state
experiments. We still consider $N$ charge qubits in a cavity, but
only two of them are resonant with the cavity and thus are
controlled by the Hamiltonian (\ref{H_eff1}). So only the two
resonant qubits are relevant and the evolution operator is given
by

\begin{equation}
\label{U_xx} U(t)=e^{i\sum\limits_{j,l=1}^2
\gamma_{jl}\sigma_j^\eta\sigma_l^\eta} e^{[\sum\limits_{j=1}^2
\alpha_j(t,\beta_j^0)\sigma_j^\eta a^\dagger-H.c.]},
\end{equation}
where $\alpha_j (t,\beta^0_j)=(g_jE_j\hbar
\delta_j)(1-e^{i\delta_j t})e^{i\beta_j^0}$ and
$\gamma_{jl} (t)=[g_j g_l E_j E_l/(\hbar^2
d_{jl}\delta_j\delta_l)][\delta_j\sin(d_{jl}t+\beta_{jl}^0)-d_{jl}\sin(\delta_j
t+\beta_{jl}^0)-\delta_l \sin\beta_{jl}^0 ]$ with
$d_{jl}=\delta_j-\delta_l$ and $\beta_{jl}^0=\beta_j^0-\beta_l^0$
when $d_{jl}\not=0$. Note the facts that $\sum_j
\alpha_j(t,0)=-\sum_j \alpha_j(t,\pi)$, and $\gamma_{jl} (t)$ is
the same for $\beta_j^0=\beta_l^0$. Therefore, the influence of
the cavity mode can also be eliminated by using two operators in
succession: two qubits evolve with $\beta_j^0=\beta_l^0=0$ for a
first period $\tau/2$ and then with $\beta_j^0=\beta_l^0=\pi$ for
the second period $\tau/2$. Moreover, the controlled phase gate
$\text{diag}(1,1,1,-1)$ is obtained when $\gamma_{jl}=-\pi$ after
performing one-qubit operations \cite{Zheng}.

Remarkably, following the approach addressed in
Ref.\cite{Zhu_PRL2003}, it is straightforward to find that the
phase $\gamma$ in the operators (\ref{U_x}-\ref{U_xx})  satisfies
the relation $\gamma=-\gamma_g$ $(\gamma_d=-2\gamma_g)$, where
$\gamma_g$ is the geometric phase and $\gamma_d$ is the dynamic
phase accumulated in the evolution. Thus $\gamma$ is an
unconventional geometric phase shift, which consists of both a
geometric component and a nonzero dynamic component, but  still
depends only on global geometric features. Because of this, the
high-fidelity of the gates in the present system may be
experimentally achieved. Note that, a recent experiment on trapped
ions demonstrated that the operation realized by unconventional
geometric phases possesses the high-fidelity\cite{Leibfried},
which really benefits from the geometric features: the phase is
determined only by the path area, not on the exact starting state
distributions, path shape, orientation in phase space, or the
passage rate to traverse the closed path. Moreover, this operation
is robust to small noncyclic
perturbations\cite{Zhu_PRL2003,Sjoqvist}.

At this stage, we  illustrate how to produce the GHZ state in this
system. Note that the operator (\ref{U_x}) is independent on the
number of qubits, by choosing $ \gamma =\pi/2$ and an initial
state $|\Psi^i\rangle=|00\cdots 0\rangle$, the final state
$|\Psi^f\rangle=e^{i\frac{\pi}{2}J_x^2}|\Psi^i\rangle$ is found to
be a GHZ state given by\cite{Sorensen2000}
\begin{equation}\label{GHZ}
|\Psi^f\rangle=\frac{1}{\sqrt{2}}[e^{-i\frac{\pi}{4}}|00\cdots
0\rangle+e^{i\pi(\frac{1}{4}+\frac{N}{2})}|11\cdots
1\rangle],\end{equation} when $N$ is even. Although the operator
(\ref{U_x}) does not directly produce the maximally entangled
state described in Eq.(\ref{GHZ}) for odd $N$,   it suffices to
get GHZ state by applying the unitary operator $U=\exp(-i\pi
J_x/2)$ in addition to application of Eq.(\ref{U_x}). Starting
from Eq (\ref{H_int1}), it is not difficult to see that  the
operation $U$ can be  implemented in the system by just choosing
$(\bar{n}_j=1/2,\phi_j^{-}=0, \phi_j^{+} \neq 0).$ To see the
feasibility of the present scheme with current technology, we now
use typical values of physical parameters in the system to
estimate  the operation time $\tau$. We have $\gamma \sim g^2
E^2\tau/\hbar^2\delta$. Therefore, $\tau \sim \pi\hbar^2\delta/2
g^2 E^2$ for $\gamma=\pi/2$, which is about $20 ns$ for $E=40\mu
ev$\cite{Nakamura}, $\hbar\omega_c=30\mu ev$,
$\delta=\omega_c/10$, and $g=10^{-2}$\cite{Yang,Girvin}. Thus
$\tau$ is significantly less than the decoherence time of qubit
($\sim 0.5\mu s$) without the protection of the cavity\cite{Vion},
and is much less than the photon lifetime of the cavity mode
$\tau_c=Q/\omega_c=22\mu s$, as the quality factor of the cavity
$Q=10^6$ was reported experimentally\cite{Day}. In this case the
voltage $V_0$ required to generate the external flux is
$9\omega_c\phi_0/20\pi \sim 13\ \mu V$, a typical voltage for
manipulating Josephson junctions.

Another useful application of the present geometric approach is to
construct QECC in this solid-state system. The operator
(\ref{U_x}) can also be used for QECC\cite{You}. For example, the
Shor coding
\begin{equation} \label{Shor}
\alpha[\frac{1}{\sqrt{2}}(|000\rangle-|111\rangle)]^{\otimes 3}
+\beta[\frac{1}{\sqrt{2}}(|000\rangle+|111\rangle)]^{\otimes 3}
\end{equation}
with $|\alpha|^2+|\beta|^2=1$ is achieved by using the operator
(\ref{U_x}) plus single qubit measurement or using the operator
(\ref{U_x}) and a coupling $\sigma_z\sigma_z$, which arises
naturally in charge qubits coupled through capacitors. Therefore,
the approach proposed here provides a geometric way to efficiently
construct this kind of essential codes in quantum information.
A significant result here is that
implementation of a universal set of quantum gates as well as
construction of QECC can be based on the same set of geometric
quantum operators with intrinsic fault tolerant features and then
may simplify the experimental operations for quantum computation.


This work was supported by the European Union project TOPQIP
(contract IST-2001-39215), the NSFC under Grant Nos. 10429401,
10204008 and 10334090, the NSF of Guangdong under Grant No.021088,
Program for NCET, the RGC grant of Hong Kong
(HKU7114/02P), and the URC fund of HKU.

\end{document}